\documentclass[paperpaper,twocolumn,english,superscriptaddress,aps,prx,floatfix,amsfonts,amssymb]{revtex4-1}

\usepackage{epsfig}
\usepackage{graphicx}
\usepackage{dcolumn}
\usepackage{bm}

\newcommand{\EQ}{\begin{equation}}
\newcommand{\EN}{\end{equation}}
\newcommand{\ea}{\end{eqnarray}}
\newcommand{\ba}{\begin{eqnarray}}

\newcommand{\bear}{\begin{eqnarray}}
\newcommand{\ear}{\end{eqnarray}}

\begin{document}
\title{Finite-temperature transport in the gapped spin-${1\over 2}$ XXZ chain
and one-dimensional lattice spinless fermion model}


\author{J. M. P. Carmelo}
\affiliation{Center of Physics of University of Minho and University of Porto, LaPMET, P-4169-007 Oporto, Portugal}
\affiliation{CeFEMA, Instituto Superior T\'ecnico, Universidade de Lisboa, LaPMET, Av. Rovisco Pais, P-1049-001 Lisboa, Portugal}

\author{P. D. Sacramento}
\affiliation{CeFEMA, Instituto Superior T\'ecnico, Universidade de Lisboa, LaPMET, Av. Rovisco Pais, P-1049-001 Lisboa, Portugal}



\begin{abstract}
It is well established that at zero magnetic field, $h=0$, the spin diffusion constant of
the spin-${1\over 2}$ XXZ chain is for anisotropy $\Delta >1$ finite for both low and high temperatures, 
implying that the type of spin transport is normal diffusive. Although it is expected that this holds for all finite temperatures, $T>0$,
the calculation of the spin diffusion constant for all $T$ is typically intractable. Here we consider a
class of energy eigenstates that exist both for anisotropies $\Delta =1$ and $\Delta >1$.
We show that at the isotropic point their contributions are behind the diffusion constant being infinite, 
spin transport being anomalous superdiffusive for $T>0$. That for $\Delta >1$ such states do not contribute to the 
diffusion constant is shown to imply it is finite, spin transport being normal diffusive for $T>0$.
By combining the connection through a Jordan-Wigner transformation of the spin-${1\over 2}$ XXZ chain
to the one-dimensional (1D) lattice spinless fermion model 
at zero chemical potential, $\mu =0$, for $V/J \geq 1$ with its Bethe-ansatz solution,
where $V$ is the nearest-neighbor Coulomb repulsion 
and $J = 2t$ is twice the hopping integral $t$, in this paper we also address the issue of the $T>0$ 
charge transport of that model at $\mu =0$. It is found to be anomalous superdiffusive for $V/J=1$
and normal diffusive for $V/J>1$. Our results thus
open the door to a key advance in the understanding for {\it all} finite temperatures $T>0$ of the 
spin transport in the spin-${1\over 2}$ XXZ chain at $h=0$ for anisotropy $\Delta \geq 1$ and of 
the charge transport in the 1D lattice spinless fermion model at $\mu =0$ for $V/J \geq 1$.
\end{abstract}
\maketitle

\section{Introduction}
\label{SECI}

There has been significant recent interest in the finite temperature $T>0$ spin transport properties of the spin-${1\over 2}$ XXZ 
chain at zero magnetic field, $h=0$, for anisotropy $\Delta \geq 1$ 
\cite{Carmelo_25,Nardis_23,Ye_22,Ilievski_21,Nardis_20,Gopalakrishnan_19,Ljubotina_19,Nardis_19,Medenjak_17,Ljubotina_17,Steinigeweg_11}.
That issue is closely related to the $T>0$ charge transport properties of the one-dimensional (1D) lattice spinless fermion model 
at zero chemical potential, $\mu = 0$, for $V/J \geq 1$ where $V$ is the nearest-neighbor Coulomb repulsion and $J=2t$ is twice the hopping
integral $t$\cite{Peres_99,Gu_02,Pereira_09,Gebhard_22}. There is indeed a connection of the spin-${1\over 2}$ XXZ 
chain to that charge model by means of a Jordan-Wigner transformation \cite{Jordan_28,Cloizeaux_66,Yang_66}, 
which can be combined with the latter model Bethe-ansatz solution \cite{Peres_99,Gu_02}.

The real part of the spin-${1\over 2}$ XXZ chain's spin conductivity ($\alpha =s$) and 1D lattice spinless fermion model's 
charge conductivity ($\alpha = c$) is for finite temperatures given by,
\begin{equation}
\sigma_{\alpha} (\omega,T) = 2\pi D_{\alpha}^z (T)\,\delta (\omega) + \sigma_{\alpha,{\rm reg}} (\omega,T) \, .
\label{sigma}
\end{equation}
When the stiffness $D_{\alpha}^z (T)$ in its singular part is finite, 
the dominant spin ($\alpha =s$) or charge ($\alpha =c$) transport is ballistic. There is strong evidence that in 
the thermodynamic limit the spin stiffness of the spin-${1\over 2}$ XXZ 
chain for anisotropy $\Delta \geq 1$ vanishes at $h=0$ for all temperatures $T>0$ \cite{Carmelo_25,Nardis_19}. 
Consistently, the same applies to the charge stiffness of the 1D lattice spinless fermion model at $\mu =0$ for $V/J \geq 1$
and $T>0$ \cite{Peres_99}.

The type of nonballistic spin or charge transport is determined by the value of the corresponding
diffusion constant $D_{\alpha} (T)$ associated with the regular part of the conductivity $\sigma_{\alpha,{\rm reg}} (\omega,T)$ in Eq. (\ref{sigma}):
It is normal diffusive when $D_{\alpha} (T)$ is finite and anomalous superdiffusive when $D_{\alpha} (T) = \infty$.

Due to the closed relation between the two models, we can study the issue of the type of $T>0$ nonballistic transport
in one of them, which under the Jordan-Wigner transformation combined with the Bethe-ansatz solution
provides information on the type of $T>0$ nonballistic transport in the other model. 

The studies of this paper focus mainly on the type of $T>0$ nonballistic 
spin transport of the spin-${1\over 2}$ XXZ chain at $h=0$ for anisotropy $\Delta \geq 1$. 
In Sec. \ref{SECV} we address and discuss the issue of the type of $T>0$ nonballistic 
charge transport of the 1D lattice spinless fermion model at $\mu =0$ for $V/J\geq 1$.

Hydrodynamic theory and Kardar-Parisi-Zhang (KPZ) scaling studies have found that the
spin transport in the isotropic spin-${1\over 2}$ XXX chain at $h=0$ is for {\it all} temperatures $T>0$ anomalous 
superdiffusive \cite{Ljubotina_19,Nardis_20,Nardis_23}. 

The KPZ scaling refers to a universality class that involves a broad range of classical stochastic 
growth models that exhibit similar scaling behavior to the original KPZ equation \cite{Kardar_86,Krug_97,Kriecherbauer_10,Corwin_12}. 
It has also been shown to describe high-temperature $T\rightarrow\infty$ and thus classical 
dynamics of certain many-body systems near equilibrium \cite{Ljubotina_19}.

An interesting and somehow unexpected result is that KPZ scaling prevails for all finite temperatures $T>0$ in the case 
of the dynamical spin structure factor of the SU(2) symmetrical spin-${1\over 2}$ XXX chain,
which was found to be exactly described by the KPZ correlation function \cite{Ljubotina_19,Nardis_20,Nardis_23}.
Within the use of a kinetic theory of transport, the KPZ coupling strength for that chain was computed as a 
function of temperature, directly from microscopics \cite{Nardis_20}.

The integrable spin-${1\over 2}$ XXX
chain has time-reversal and parity symmetries that are absent from the
KPZ equation. Those force higher-order spin fluctuations deviate from standard KPZ predictions. 
A nonlinear fluctuating hydrodynamic theory consisting of two coupled stochastic modes,
was used to explain the emergence of anomalous spin dynamics in the spin-${1\over 2}$ XXX
chain at $h=0$ \cite{Nardis_23}. Indeed, it predicts KPZ scaling for the spin structure factor but with a symmetric, 
quasi-Gaussian, distribution of spin fluctuations. Whether KPZ dynamics occurs in all integrable spin chains 
with non-Abelian symmetry is an interesting issue under debate \cite{Ye_22,Ilievski_21}.

We focus our studies mostly on the finite-temperature spin transport in the spin-${1\over 2}$ XXZ chain 
at $h=0$ for anisotropy $\Delta >1$. It is well established that for that spin chain at $h=0$
the spin diffusion constant associated with the regular part of the spin conductivity $\sigma_{s,{\rm reg}} (\omega,T)$
[Eq. (\ref{sigma}) for $\alpha =s$] is finite both for very low temperatures \cite{Carmelo_25} and high temperatures $T \in [J/k_B,\infty]$
\cite{Carmelo_25,Gopalakrishnan_19,Steinigeweg_11}. The spin diffusion constant reaches at low temperatures 
larger values than at at high temperatures. At both very low temperatures and high
temperatures $T \in [J/k_B,\infty]$ it decreases upon increasing $T$ \cite{Carmelo_25}. Hence it is
expected that the spin diffusion constant has finite values for all temperatures, $T>0$ \cite{Carmelo_25,Ljubotina_17}.

However, the calculation of the spin diffusion constant for all $T$ is typically intractable. In this paper we consider a
class of energy eigenstates that exist both for anisotropies $\Delta =1$ and $\Delta >1$. We show why at the isotropic 
point such states are behind the spin diffusion constant being infinite for all finite $T$. In contrast, for $\Delta >1$ they do not contribute 
to that diffusion constant at finite temperatures $T>0$. And this property is shown in this paper to justify why for $\Delta >1$ the spin diffusion 
constant is finite at $h=0$ for all finite temperatures $T>0$. Hence $T>0$ spin transport is anomalous superdiffusive and
normal diffusive for $\Delta =1$ and $\Delta >1$, respectively.

Similar results are obtained for the 1D lattice spinless fermion model at $\mu =0$
for $T>0$, the charge diffusion constant of which is found to be infinite at $V/J =1$ and finite for $V/J >1$.

The paper is organized as follows. An exact paired and unpaired physical-spin representation 
of the spin-${1\over 2}$ XXZ chain for anisotropy $\Delta \geq 1$ \cite{Carmelo_25,Carmelo_15}
is the subject of Sec. \ref{SECII}. In Sec. \ref{SECIII} the 
spin diffusion constant is expressed in terms of the spin elementary 
currents carried by the spin carriers, which at $h=0$ are the unpaired physical spins in the model's global $q$-spin continuous SU$_q$(2) symmetry 
multiplet configurations of $S_q >0$ energy eigenstates. Here $S_q$ is the $q$-spin \cite{Carmelo_25,Pasquier_90}. At $\Delta =1$,
they are the unpaired physical spins in the SU (2) symmetry multiplet configurations of finite-spin $S >0$ energy 
eigenstates \cite{Carmelo_15}. The contributions to the spin diffusion constant from states we call of classes (A) and (B), 
respectively, is the issue addressed in Sec. \ref{SECIV}. 
Sec. \ref{SECV} is devoted to the study of the $T>0$ charge transport in the 1D lattice spinless fermion model at $\mu =0$
for $V/J \geq 1$. The concluding remarks are presented in Sec. \ref{SECVI}. Some side results needed for the studies of 
this paper on the states of class (B) of the spin-${1\over 2}$ XXZ chain at $h=0$ for $\Delta \geq 1$ are presented in the Appendix.
 
\section{The $\Delta >1$ spin-${1\over 2}$ XXZ chain in terms of paired and unpaired physical spins}
\label{SECII}  

For spin anisotropy $\Delta=\cosh\eta \geq 1$ and thus $\eta \geq 0$, exchange integral $J$, and lattice length 
$L\rightarrow\infty$ for $N/L$ finite, the Hamiltonian of the antiferromagnetic spin-${1\over 2}$ XXZ chain at 
$h = 0$ is given by,
\begin{equation}
\hat{H} = J\sum_{j=1}^{N}\left({\hat{S}}_j^x{\hat{S}}_{j+1}^x + {\hat{S}}_j^y{\hat{S}}_{j+1}^y + 
\Delta\,{\hat{S}}_j^z{\hat{S}}_{j+1}^z\right) \, .
\label{HD1}
\end{equation}
Here $\hat{\vec{S}}_{j}$ is the spin-${1\over 2}$ operator at site $j=1,...,N$ with components $\hat{S}_j^{x,y,z}$.
This Hamiltonian describes the correlations of
$N=\sum_{\sigma =\uparrow,\downarrow}N_{\sigma}$ physical spins $1/2$. 
We use natural units in which the Planck constant 
and the lattice spacing are equal to 1, so that $N=L$.

For anisotropy $\Delta > 1$ the spin chain model, Eq. (\ref{HD1}), has a global $q$-spin continuous SU$_q$(2) symmetry
the irreducible representations of which are isomorphic to those of the $\Delta = 1$ SU (2) symmetry
\cite{Carmelo_25,Pasquier_90}. For $\Delta >1$, the spin projection $S^z$ remains a good quantum 
number whereas spin $S$ is not. It is replaced by the $q$-spin $S_q$ in the eigenvalues of the Casimir 
generator of the continuous SU$_q$(2) symmetry \cite{Carmelo_25,Pasquier_90}. The values of $q$-spin $S_q$ are exactly the 
same for anisotropy $\Delta >1$ as those of spin $S$ for $\Delta =1$. This includes their relation to the values 
of $S^z$. Hence {\it singlet} and {\it multiplet} refer in this paper to physical-spins configurations with 
zero and finite $q$-spin $S_q$, respectively. For $\Delta = 1$ such configurations refer to the usual spin $S$.

Representations in terms of spinons and similar quasi-particles such as psinons 
and antipsinons \cite{Karbach_02} have been widely used to successfully describe the 
static and dynamical properties of both spin-chain models and the physics of the materials they represent. 
Hence such representations became the paradigm of the spin-chains physics. 

However, at $h=0$ and finite temperatures $T>0$, the contributions to spin transport involve
a huge number of energy eigenstates many of which are outside the subspaces where
such representations are valid. Indeed, many of such states are generated in the thermodynamic limit 
from ground states by an infinite number of elementary microscopic processes. This renders the usual spinon and alike 
representations unsuitable to handle that very complex finite-temperature quantum problem. 
On the other hand, the paired and unpaired physical spins representation \cite{Carmelo_25,Carmelo_15} 
used in the studies of this paper refers to the model's whole Hilbert space.

The $q$-spin continuous SU$_q$(2) symmetry for $\Delta >1$ \cite{Carmelo_25,Pasquier_90} and
the spin SU(2) symmetry at $\Delta = 1$ \cite{Carmelo_15} impose that for 
{\it all} energy eigenstates the number $N$ of physical spins described by the Hamiltonian, Eq. (\ref{HD1}), can 
be written as $N = M + {\cal{M}}$. 

Here $M = M_{+1/2}+M_{-1/2} = 2S_q$ 
where $S_q = S$ for $\Delta =1$ and $M_{\pm 1/2}= S_q \pm S^z$ is the number of  {\it unpaired physical spins} of projection $\pm 1/2$ in 
the multiplet configuration of $S_q>0$ energy eigenstates. On the other hand, ${\cal{M}} = {\cal{M}}_{+1/2} + {\cal{M}}_{-1/2}$ 
where ${\cal{M}}_{+ 1/2} = {\cal{M}}_{- 1/2} = N/2 - S_q$ and thus ${\cal{M}} = N - 2S_q$ is that of {\it paired physical spins} 
in singlet configurations of such states.

For each energy eigenstate, the paired physical spins are contained in a number $N/2-S_{q}$ of singlet pairs where
$S_q = S$ at $\Delta =1$. Such singlet pairs are distributed over singlet configurations, each containing a number $n=1,...,\infty$ of pairs, which we call
$n$-pairs and the number of which we denote by $N_n$. Hence the number of singlet pairs reads $N/2-S_{q} =  \sum_{n=1}^{\infty} n N_n$,
so that ${\cal{M}}  = \sum_{n=1}^{\infty} 2n N_n$ and ${\cal{M}}_{+ 1/2} = {\cal{M}}_{- 1/2} = \sum_{n=1}^{\infty} n N_n$.

The number $N$ of physical spins can then be written as,
\begin{equation}
N = M + {\cal{M}} = 2S_q + \sum_{n=1}^{\infty}2n N_{n} \, .
\label{N}
\end{equation}

There is an $n$-band for each set of $N_n$ $n$-pairs with fixed $n$ value.
For both $n=1$ and $n>1$, the number $L_n$ of $j=1,...,L_n$ discrete momentum 
values $q_j$ in each $n$-band is given by \cite{Carmelo_25},
\begin{equation}
L_ n = N_n + N_n^h \hspace{0.20cm}{\rm where}\hspace{0.20cm}
N_n^h = 2S_q + \sum_{n'=n+1}^{\infty}2(n'-n)N_{n'} \, .
\label{LnNnh}
\end{equation}
Here $N_n$ is the number of occupied $q_j$'s and thus of $n$-pairs and $N^h_{n}$ 
is that of unoccupied $q_j$'s. We call them $n$-holes. 

The Bethe-ansatz quantum numbers $I_j^n$ \cite{Gaudin_71} are actually
the $n$-band shifted discrete $n$-band momentum values $q_j - q_n^{\Delta} = {2\pi\over N}I_j^n$
in units of ${2\pi\over N}$. Here $q_n^{\Delta}$ has limiting values given in Eq. (\ref{qDelta}) 
of the Appendix. It vanishes for $\Delta \rightarrow 1$ and for $\Delta >1$ 
vanishes in the thermodynamic limit for ground states and excited states generated from them by a finite number 
of $n$-band processes.

The Bethe-ansatz quantum numbers read as $I_j^n = 0,\pm 1,...,\pm {L_n -1\over 2}$ for $L_n$ odd
and $I_j^n = \pm 1/2,\pm 3/2,...,\pm {L_n -1\over 2}$ for $L_n$ even. 
The set $\{q_j\}$ of discrete $n$-band $q_j$'s have Pauli-like 
occupancies: The corresponding $n$-band momentum distributions read as $N_n (q_j) = 1$ and 
$N_n (q_j) = 0$ for occupied and unoccupied $q_j$'s, respectively.

Since $q_{j+1}-q_j = {2\pi\over N}$, the $n$-band shifted discrete momentum values $q_j - q_n^{\Delta}$'s
can in the thermodynamic limit be described by continuous $n$-band shifted momentum variables $q - q_n^{\Delta} \in [q_n^-,q_n^+]$
the limiting values $q_n^{\pm}$ of which are given in Eq. (\ref{qqq}) of the Appendix.

As shown in Ref. \onlinecite{Carmelo_25}, in the presence of a uniform vector potential $\Phi = \Phi_{\uparrow} = -\Phi_{\downarrow}$
(twisted boundary conditions), only the $M = M_{+1/2}+M_{1/2} = 2S_q$ unpaired physical spins couple
to it and contribute to spin transport, so that they are the {\it spin carriers}. In that reference it was found that the coupling to spin
of the number $M_{\pm 1/2}= S_q \pm S^z$ of unpaired physical spins of opposite projection $\pm 1/2$ in the multiplet
configuration of the $S_q>0$ energy eigenstates also has opposite sign. 

Importantly, each such a spin carrier carries a spin elementary current $j_{s,\pm 1/2}$ the general expression of which is
given in Ref. \onlinecite{Carmelo_25} for all $S_q >0$ energy eigenstates. The expectation values of the
$z$ component of the spin-current operator,
\begin{equation}
\hat{J}^z = i\,{J\over 2}\sum_{j=1}^{N}(\hat{S}_j^+\hat{S}_{j+1}^- - \hat{S}_{j+1}^+\hat{S}_j^-)  \, ,
\label{s-current}
\end{equation}
can be expressed in terms of such spin elementary currents \cite{Carmelo_25}.

As confirmed in the following, these spin elementary currents play a key role in 
spin transport at finite temperatures. 

\section{The spin diffusion constant in terms of spin elementary currents}
\label{SECIII}  

From manipulations of the Kubo formula and Einstein relation, 
the spin-diffusion constant $D_s (T)$ associated with the regular part of the 
spin conductivity $\sigma_{s,{\rm reg}} (\omega,T)$, Eq. (\ref{sigma}) for $\alpha =s$, can
be expressed in terms of the spin elementary currents $j_{s,\pm 1/2}$ carried by the
spin carriers as \cite{Carmelo_25},
\begin{equation}
D_s (T) = C_s (T)\,\Pi_s (T) = C_s (T)\,N\,\langle\vert j_{s,\pm 1/2}\vert^2\rangle_{T} \, .
\label{DproptoOmega}
\end{equation}

Here the thermal expectation value reads,
\begin{eqnarray}
\langle\vert j_{s,\pm 1/2}\vert^2\rangle_{T} & =&  {\Pi_s (T)\over N} 
\nonumber \\
& = & \sum_{S_{q}=1}^{N/2}\sum_{S^z=-S_{q}}^{S_{q}}\sum_{l_{\rm r}^{\eta}} 
p_{l_{\rm r}^{\eta},S_{q},S^z}\vert j_{s,\pm 1/2}\vert^2 \, ,
\label{jz2TD}
\end{eqnarray}
where $p_{l_{\rm r}^{\eta},S_q,S^z}$ are the Boltzmann weights and $l_{\rm r}^{\eta}$ refers to the set of quantum numbers 
other than $S_{q}$ and $S^z$ needed to specify an energy eigenstate.

The coefficient $C_s (T) = 1/[8 v_{s,{\rm LR}}\,\chi_s (T)\,f_{1} (T)]$ in Eq. (\ref{DproptoOmega}) is finite for $T>0$. In its  
expression, $v_{s,{\rm LR}}$ and $\chi_s (T)$ are the spin Lieb-Robinson velocity and static spin susceptibility, respectively, 
and $f_{1} (T)$ is the second derivative of the free-energy density with respect to 
$m = 2S^z/N$ at $m = 0$. The spin Lieb-Robinson velocity is the maximal velocity with which the information can travel.

Since the coefficient $C_s (T)$ is finite for $T>0$,
the form of the spin-diffusion constant expression, Eq. (\ref{DproptoOmega}), shows 
that the thermal expectation value $\langle\vert j_{s,\pm 1/2}\vert^2\rangle_{T} = \Pi_s (T)/L$, Eq. (\ref{jz2TD}), 
of the square of the spin elementary currents $j_{s,\pm 1/2}$ carried by each of the number $M = 2S_q$ of spin carriers
that populate a $S_q>0$ energy eigenstate controls the type of spin transport. Specifically, according to the following criteria one has,
\begin{eqnarray}
&& \Pi_s (T) = N\,\langle\vert j_{s,\pm 1/2}\vert^2\rangle_{T}\rightarrow
\infty \Rightarrow {\rm spin}\hspace{0.20cm}{\rm superdiffusion}
\nonumber \\
&& \Pi_s (T) = N\,\langle\vert j_{s,\pm 1/2}\vert^2\rangle_{T}
\hspace{0.20cm}{\rm finite} \Rightarrow {\rm spin}\hspace{0.20cm}{\rm diffusion} \, .
\label{noramomD}
\end{eqnarray}
\begin{figure*}
\includegraphics[width=0.495\textwidth]{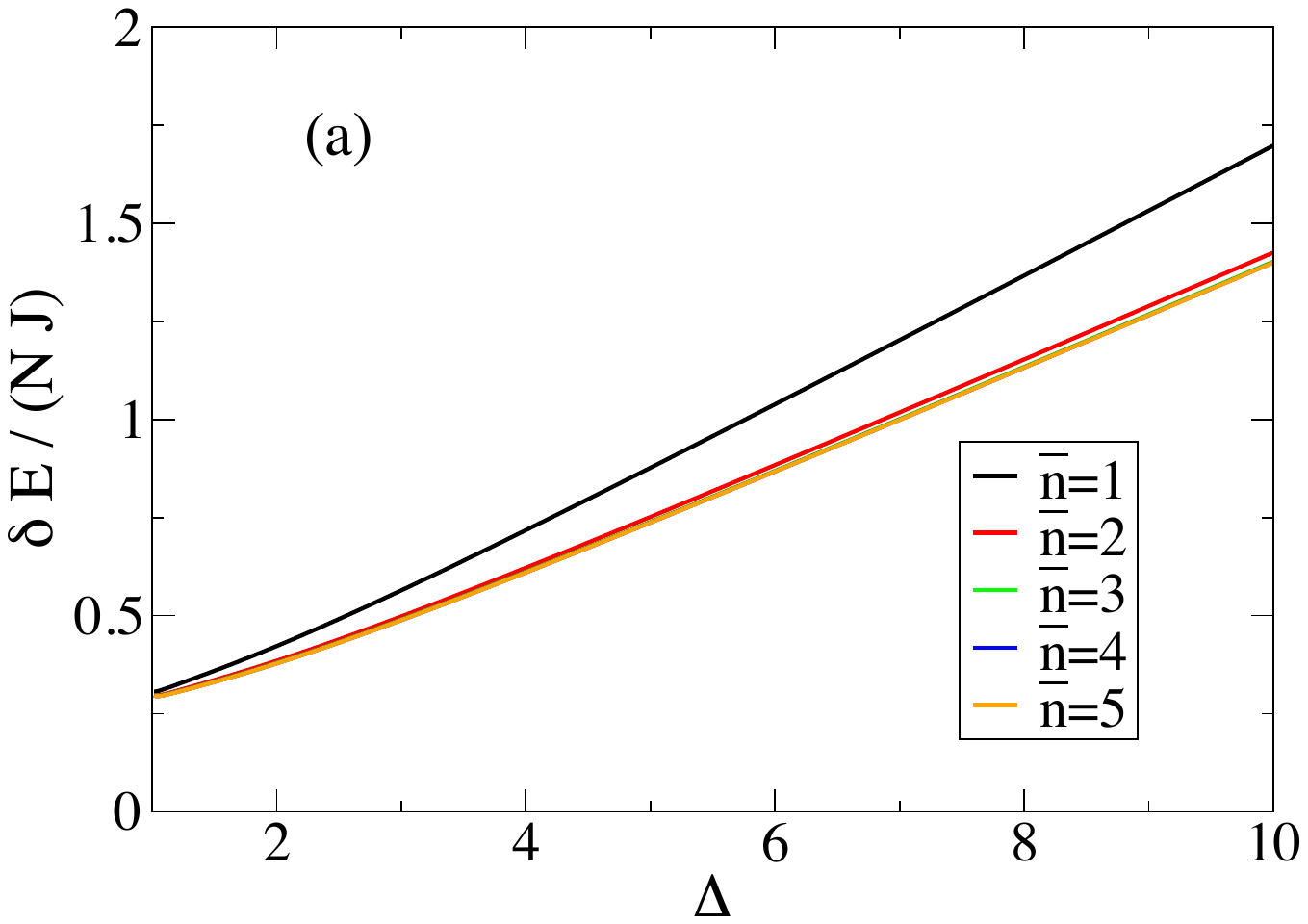}
\includegraphics[width=0.495\textwidth]{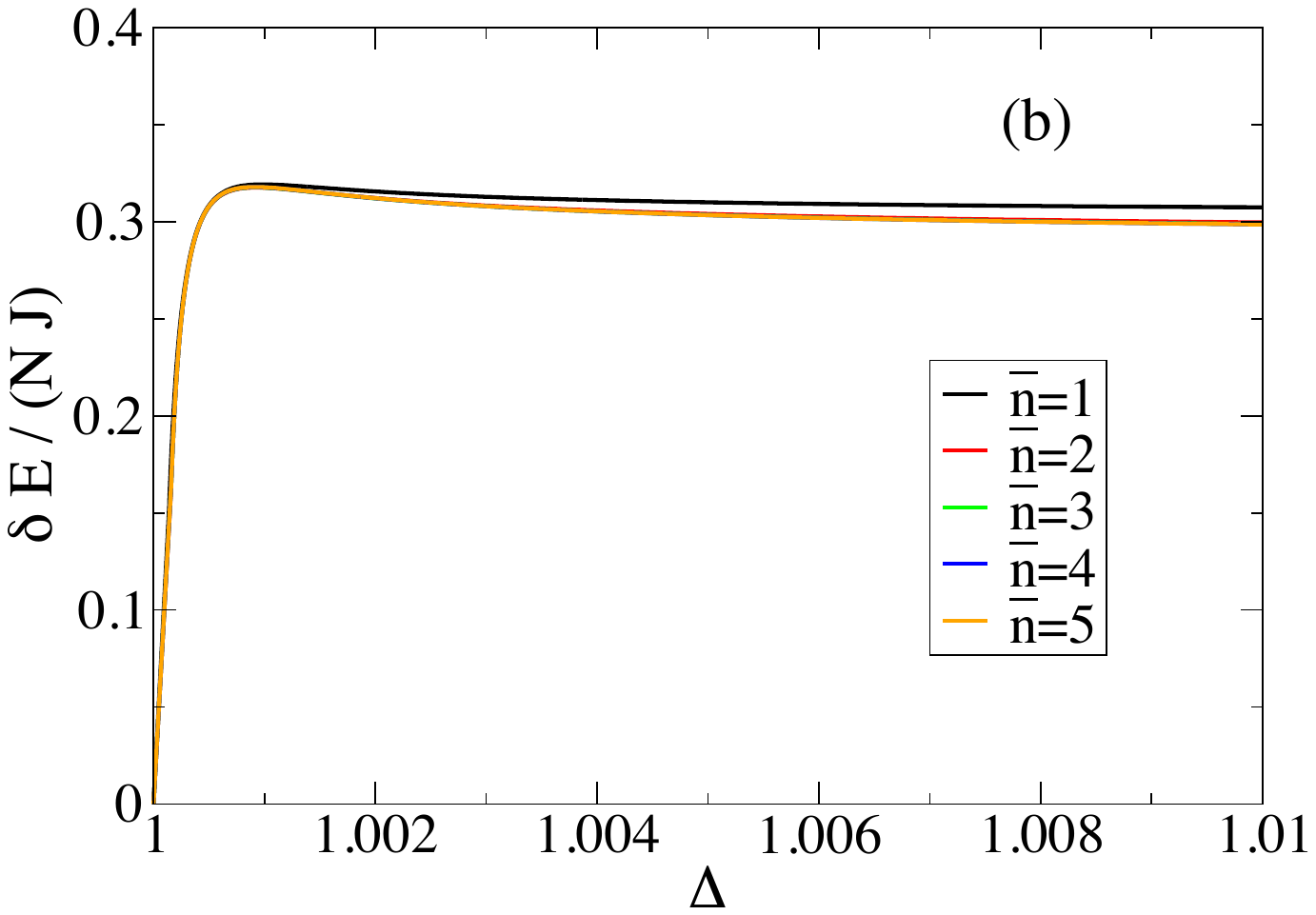}
\caption{The excitation-energy density $\delta E/(N J)$ of $\bar{n}$ states with
$\bar{n} = 1,2,3,4,5$ in units of $J$ plotted as a function of the anisotropy 
for (a) $\Delta \in ]1.01,10]$ and (b) $\Delta \in ]1,1.01]$.}
\label{figure1XXZ_25}
\end{figure*}

Analytical expressions of the spin-diffusion constant have been obtained both for very low temperatures \cite{Carmelo_25} and 
high temperatures $T \in [J/k_B,\infty]$ \cite{Carmelo_25,Gopalakrishnan_19,Steinigeweg_11}.
However, the calculation for all temperatures of the spin-diffusion constant's general expressions given in Eq. (\ref{DproptoOmega}),
Eq. (6.38) of Ref. \onlinecite{Nardis_19}, and Eq. (5) of Ref. \onlinecite{Medenjak_17} is typically intractable.  

Useful physical information can though be extracted from the expression of the spin-diffusion constant, Eq. (\ref{DproptoOmega}).
Indeed, $\vert j_{s,\pm 1/2}\vert^2$ in Eq. (\ref{jz2TD}) does not depend on the temperature and the 
temperature dependence of the Boltzmann weights $p_{l_{\rm r}^{\eta},S_{q},S^z}$ in that equation
and of the static spin susceptibility $\chi_s (T)$ and second derivative of the free-energy density $f_{1} (T)$
in the expression of the coefficient $C_s (T) = 1/[8 v_{s,{\rm LR}}\,\chi_s (T)\,f_{1} (T)]$ 
appearing in Eq. (\ref{DproptoOmega}) is monotonous in the case of the spin-${1\over 2}$ XXZ chain for $T>0$ and $\Delta >1$. This implies
that the temperature dependence of the spin diffusion constant is also monotonous. 
This can be explicitly confirmed in the limits of both low and high temperature. 

In the limit of very low temperature the spin gap is for $\Delta >1$ given by $E_{\rm gap} = {2J\over\pi}\sinh \eta\,K (u_{\eta})\,(1 - u_{\eta}^2)^{1/2}$
where $K (u_{\eta})$ is the complete elliptic integral and the $\eta$-dependence of the parameter 
$u_{\eta}$ is defined by the relation $\eta = \pi K' (u_{\eta}) /K (u_{\eta})$ where $K' (u_{\eta}) = K ((1 - u_{\eta}^2)^{1/2})$. 
The spin carriers of the excited states are $S^z=\pm 1$ triplet pairs that involve two unpaired physical 
spins with the same projection $\pm 1/2$ and have transport mass $M_t$, the inverse of which is
given by $1/M_t = {2J\over\pi}\sinh \eta\,K (u_{\eta})\,u_{\eta}^2 (1 - u_{\eta}^2)^{-1/2}$.
In that limit of very low temperature the spin diffusion constant is a monotonous decreasing function of the
temperature given by $D_s (T) = {e^{E_{\rm gap}/K_B T}\over 8M_t}$ \cite{Carmelo_25}. 

While $\lim_{T\rightarrow 0}D_s (T) = \infty$ for $\Delta >1$, this result does not apply at $T=0$. Indeed,
the exponential factor $e^{E_{\rm gap}/K_B T}$ in the spin diffusion constant expression stems from
the average distance $x_p = (4M_t k_B T)^{-1/2}\,e^{E_{\rm gap}/k_B T}$ traveled by one triplet pair before it interacts 
with other triplet pairs. On decreasing the temperature, the number of triplet pairs also decreases and
thus the average distance $x_p$ increases. At $T=0$ there are no triplet pairs, so that $x_p$ 
ceases to have physical meaning and the expression $D_s (T) = {e^{E_{\rm gap}/K_B T}\over 8M_t}$ 
does not apply. 

In the opposite limit of very high temperatures the spin diffusion constant is also a monotonous
decreasing function of the temperature. It is plotted in Fig. 4 (a) of Ref. \onlinecite{Carmelo_25}
as a function of $\beta J = J/(k_B T) \in [0,1]$ for several values of the anisotropy. 

The results for very low and high temperatures suggest the possibility of the spin diffusion
constant being a monotonous decreasing function of the temperature, decreasing on increasing $T$ from $\lim_{T\rightarrow 0}D_s (T) = \infty$
to its minimal value reached in the $T\rightarrow \infty$ limit \cite{Carmelo_25,Gopalakrishnan_19},
\begin{eqnarray}
&& \lim_{T\rightarrow\infty} D_s (T) = {4J\over 9\pi}\sqrt{\Delta^2 - 1}\,\sum_{n=1}^{\infty}(n+1)
\nonumber \\
&& \times \Bigl({n+2\over (\Delta + \sqrt{\Delta^2 -1})^n - (\Delta - \sqrt{\Delta^2 -1})^n}
\nonumber \\
&& - {n\over (\Delta + \sqrt{\Delta^2 -1})^{n+2} - (\Delta - \sqrt{\Delta^2 -1})^{n+2}}\Bigr) \, .
\label{DsTDeltainfty}
\end{eqnarray}

Our results are valid in the thermodynamic limit, $N\rightarrow\infty$. This ensures that the spin-${1\over 2}$ XXZ chain 
has infinitely many conserved quantities associated with the $n=1,...,\infty$ $n$-pairs.
In that limit the imaginary part of the Bethe-ansatz rapidities has a very simple form.

As discussed below in Sec. \ref{SECIV}, $\langle\vert j_{s,\pm 1/2}\vert^2\rangle_{T}$ [Eq. (\ref{jz2TD})]
is in the thermodynamic limit of the order of $1/N$ for $\Delta >1$ and finite at $\Delta = 1$.
For a large finite system for which the $n$-pairs can have large values 
for the number $n$ of spin pairs, the physics remains nearly the same for $\Delta >1$. Indeed, 
$\langle\vert j_{s,\pm 1/2}\vert^2\rangle_{T}$ in the second expression of Eq. (\ref{noramomD}) 
remains of the order of $1/N$, so that $\Pi_s (T) = N\,\langle\vert j_{s,\pm 1/2}\vert^2\rangle_{T}$ remains
finite. On the other hand, it is expected that at the isotropic point, $\Delta =1$, superdiffusion becomes 
diffusion at large finite values of $N$. Indeed, $\langle\vert j_{s,\pm 1/2}\vert^2\rangle_{T}$ remains
finite and thus $\Pi_s (T) = N\,\langle\vert j_{s,\pm 1/2}\vert^2\rangle_{T}$ 
in the first expression of Eq. (\ref{noramomD}) becomes large yet finite.

\section{Contributions to the spin diffusion constant from states of classes (A) and (B), respectively}
\label{SECIV}  

We can divide the $S_q>0$ energy eigenstates into those of class (A) for which $\vert j_{s,\pm 1/2}\vert^2$ is of the order of $1/N$ and vanishes in the thermodynamic limit, $N\rightarrow\infty$, and those of class (B) for which $\vert j_{s,\pm 1/2}\vert^2$ is finite. 
(For anisotropy $\Delta >1$ there are no energy eigenstates for which $\vert j_{s,\pm 1/2}\vert^2$ is infinite \cite{Carmelo_25}.)

The form of the spin diffusion-constant expression, $D_s (T) = C_s (T)\,N\,\langle\vert j_{s,\pm 1/2}\vert^2\rangle_{T}$, Eq. (\ref{DproptoOmega}), shows that 
the finite density of energy eigenstates of class (B) for which $\vert j_{s,\pm 1/2}\vert^2$ is finite
gives in the $N\rightarrow\infty$ thermodynamic limit an infinite contribution to the spin
diffusion constant at $k_B T$ values corresponding to such states excitations energies associated with their energy eigenvalues.

On the other hand, class (A) states provide a finite contribution to that constant.
Consistently and as justified in the following, the type of excitation-energy density spectrum of class (B) $S_q>0$ energy eigenstates 
determines whether the $T>0$ spin transport is normal diffusive or anomalous superdiffusive. 

Alternatively to the $n$-band momentum distributions $N_n (q)$, in the thermodynamic limit
we can use $n$-band rapidity distributions, 
\begin{equation}
\tilde{N}_n (\varphi) = N_n (q_n (\varphi)) \hspace{0.20cm}{\rm such}\hspace{0.20cm}{\rm that}\hspace{0.20cm}
{d q_n (\varphi)\over d\varphi} = 2\pi\sigma_{n} (\varphi) \, ,
\label{dqdvar}
\end{equation}
where the $n$-bands transformations Jacobians are the functions $2\pi\sigma_{n} (\varphi)$
that are solution of coupled integral equations given in Eqs. (A3)-(A5) of Ref. \onlinecite{Carmelo_25}.
The corresponding $n$-band momentum functions $q_n (\varphi)$ provide the values of the $n$-band momentum value
$q$ for each rapidity variable value $\varphi \in [-\pi,\pi]$. Their inverse $n$-band rapidity functions
$\varphi_n (q)$ give the $n$-band rapidity value $\varphi$ for each $n$-band momentum value 
$q \in [q_n^- + q_n^{\Delta},q_n^+ + q_n^{\Delta}]$.
Such functions are solutions of Bethe-ansatz equations \cite{Carmelo_25,Gaudin_71}
and obey the boundary conditions $q_n (\pm\pi) = q_n^{\pm} + q_n^{\Delta}$ 
and $\varphi_n (q_n^{\pm} + q_n^{\Delta}) = \pm\pi$ for all $n=1,...,\infty$ bands, respectively.

The spin elementary current's general expression obtained in Ref. \onlinecite{Carmelo_25} is expressed  
in terms of $n$-band hole distributions,
\begin{equation}
N_n^h (q) = 1 - N_n (q) \hspace{0.20cm}{\rm or}\hspace{0.20cm}
\tilde{N}_n^h (\varphi) = 1 - \tilde{N}_n (\varphi) \, .
\label{dqdvarholes}
\end{equation}

The use of such a general expression of $j_{s,\pm 1/2}$ given in Ref.
\onlinecite{Carmelo_25} in terms of the occupancies of $n$-holes of the $n$-bands with rapidity variables $\varphi \in [-\pi,\pi]$ 
justifies why the square of their absolute values are for the states of class (A) of the order of $1/N$: Most 
$n$-bands occupancy configurations that generate such energy eigenstates have both occupied and unoccupied rapidity variables
values for $\varphi <0$ and $\varphi >0$, respectively. This implies that many of the contributions to the spin elementary currents
from $\varphi <0$ and $\varphi >0$ occupancies cancel each other. 

In contrast, the states of class (B) have compact occupancies of $n$-holes and $n$-pairs with 
rapidity variables $\varphi$ of opposite sign, respectively, what provides finite contributions to the spin elementary currents. 

The finite spin elementary currents with largest absolute values have in rapidity space 
$n$-bands the $n$-holes of which have compact occupancy for $\varphi <0$ and the $n$-pairs
of which have compact occupancy for $\varphi >0$, respectively, or vice versa. 

Here we consider the states of class (B) that have the largest absolute values 
$\vert j_{s,\pm 1/2}\vert$. They are called $\bar{n}$ states and
their rapidity-variable distributions are given by \cite{Carmelo_25},
\begin{eqnarray}
\tilde{N}_{n} (\varphi) & = & 0 \hspace{0.20cm}{\rm for}\hspace{0.20cm}\varphi\in [-\pi,0]
 \hspace{0.20cm}{\rm and}\hspace{0.20cm}n=1,...,\bar{n}
\nonumber \\
& = & 1 \hspace{0.20cm}{\rm for}\hspace{0.20cm}\varphi\in [0,\pi]
 \hspace{0.20cm}{\rm and}\hspace{0.20cm}n=1,...,\bar{n}
\nonumber \\
\tilde{N}_{n} (\varphi) & = & 0  \hspace{0.20cm}{\rm for}\hspace{0.20cm}n > \bar{n} \, ,
\label{Nnstates}
\end{eqnarray}
where $\bar{n} = 1,...,\infty$. They thus have half-filled rapidity-variable occupancies for $n$-band distributions with $n$ given
by $n=1,...,\bar{n}$ and no $n$-pairs for $n>\bar{n}$. Full $n$ bands without $n$-holes and
empty $n$ bands without $n$-pairs do not contribute to spin transport.

The spin elementary currents of $\bar{n}$ states have simpler expressions
in terms of $n$-band rapidity hole distributions $\tilde{N}_n^h (\varphi)$, Eq. (\ref{dqdvarholes}).
For the $\bar{n}$ states distributions, Eq. (\ref{Nnstates}), this gives 
half-filled occupancies in rapidity variable $\varphi\in [-\pi,\pi]$ space 
for the spin elementary currents, which read as \cite{Carmelo_25},
\begin{eqnarray}
j_{s,\pm 1/2} (\bar{n}) & = & \mp {J\sinh\eta\over\pi}
\nonumber \\
& \times & \sum_{n=1}^{\bar{n}}{N\over N_n^h}
\int_{-\pi}^{0}d\varphi\,{n \sinh (n\eta)\sin \varphi\over (\cosh (n\eta) - \cos \varphi)^2} 
\nonumber \\
& = & \pm {2J\over\pi}\sum_{n=1}^{\bar{n}}{n\,\sinh\eta\over\sinh (n\eta)}{N\over N_n^h} \, .
\label{jznnRAPbarn}
\end{eqnarray}
The ratios $N/N_n^h$ have involved $\Delta$-dependent expressions given in Ref. \onlinecite{Carmelo_25}.
The related ratios $N_n/N_n^h$ are also dependent on $\Delta$, reading as $N_n/N_n^h=1$ for $\Delta\rightarrow\infty$
and decreasing on decreasing the anisotropy values. 

As discussed in the Appendix, the $\bar{n}$ states distributions, Eq. (\ref{jznnRAPbarn}), have
under the transformations, Eq. (\ref{dqdvar}), half-filled $n$-band shifted momentum $q - q_n^{\Delta}$ occupancies 
only in the $\Delta\rightarrow\infty$ limit \cite{Carmelo_25}.

It was shown in Ref. \onlinecite{Carmelo_25} that the absolute values $\vert j_{s,\pm 1/2} (\bar{n})\vert$ increase upon 
increasing $\bar{n}$. The square $\vert j_{s,\pm 1/2}(\bar{n})\vert^2$ of such spin elementary current 
absolute values is plotted in Fig. 1 of Ref. \onlinecite{Carmelo_25} for $\bar{n} = 1,2,3,4,5$ as a function of (b) 
anisotropy $\Delta$ and (d) $1/\Delta$, respectively. 

The corresponding excitation-energy density in units of $J$ of $\bar{n}$ states is for $\Delta >1$ and 
zero magnetic field given by,
\begin{eqnarray}
{\delta E\over N J} & = & \sinh\eta\,\Bigl({1\over 2} + 2 \sum_{l = 1}^{\infty}{1\over 1 + e^{2l\eta}}
\nonumber \\
& - & \sum_{n=1}^{\bar{n}}{1\over 2\pi}\int_{0}^{\pi}d\varphi\,2\pi\sigma_n (\varphi) 
{\sinh (n\,\eta)\over \cosh (n\,\eta) - \cos\varphi}\Bigr) \, ,
\nonumber \\
\label{Energy}
\end{eqnarray}
where the functions $2\pi\sigma_n (\varphi)$ are the Jacobians in Eq. (\ref{dqdvar}).
The excitation-energy density, Eq. (\ref{Energy}), is plotted in Fig. \ref{figure1XXZ_25} as a function of the anisotropy $\Delta$ for (a) 
$\Delta \in [1.01,10]$ and (b) $\Delta \in [1,1.01]$ for $\bar{n}$ states with $\bar{n} = 1,2,3,4,5$. 

Importantly, in the thermodynamic limit a finite excitation-energy density $\delta E/N J$ implies that the excitation energy $\delta E$ is
infinite. As shown in Fig. \ref{figure1XXZ_25}, for $\bar{n}$ states with $\bar{n} = 1,2,3,4,5$ the energy
density is finite for $\Delta >1$ whereas it vanishes only at $\Delta = 1$. 

This shows that the spin diffusion constant, Eq. (\ref{DproptoOmega}), has contributions from the finite density of states of class (B) with 
vanishing excitation-energy densities and thus finite excitation energies only at $\Delta = 1$. 
Indeed, the vanishing excitation-energy density $\delta E/N J=0$ at the isotropic point, $\Delta = 1$, refers to all finite values 
of the excitation energy $\delta E\in ]0,\infty]$ of $\bar{n}$ states. That at $\Delta = 1$ the quantity 
$\Pi (T)=N\,\langle\vert j_{s,\pm 1/2}\vert^2\rangle_{T}$ 
in Eq. (\ref{DproptoOmega}) is infinite in the thermodynamic limit, $N\rightarrow\infty$, is due to the contributions of
a finite density of such states of class (B) for which $\vert j_{s,\pm 1/2}\vert$ is finite and $\delta E \in ]0,\infty]$. This implies
that the spin diffusion constant of the spin-${1\over 2}$ XXX chain at $h=0$ diverges for $k_B T \in ]0,\infty]$, as also found by 
hydrodynamic theory and KPZ scaling \cite{Nardis_23,Ye_22,Ilievski_21,Nardis_20,Gopalakrishnan_19,Ljubotina_19,Nardis_19,Medenjak_17,Ljubotina_17}.

On the other hand, for $\Delta > 1$ the excitation-energy density $\delta E/N J$ is {\it always} finite, so that the excitation
energy $\delta  E$ of the $\bar{n} = 1,2,3,4,5$ states explicitly considered here is infinite. Since, as discussed in
the Appendix, this applies to all states of class (B), it follows that
for {\it all} finite temperatures $T>0$ such that $k_B T < \delta  E = \infty$ the 
spin diffusion constant, Eq. (\ref{DproptoOmega}), of the spin-${1\over 2}$ XXZ chain at $h=0$, Eq. (\ref{DproptoOmega}), is finite. 

Indeed, for $\Delta > 1$ there are no states of class (B) with finite excitation energy 
to render $\Pi (T)=N\,\langle\vert j_{s,\pm 1/2}\vert^2\rangle_{T}$  
infinite at finite $k_B T$ in $D_s (T) = C_s (T)\,\Pi_s  (T)$. In addition, $\vert j_{s,\pm 1/2}\vert^2 \propto 1/N$ 
for the states of class (A), so that they provide a finite contribution to the spin diffusion constant. 

As reported in the Appendix, we considered all different types of states of class (B) with finite occupancies
in $n$ bands for $n=1,...,\bar{n}$ where $\bar{n}=1,...,\infty$.
Importantly, all such states have excitation-energy density spectra in function of anisotropy $\Delta$ similar to those 
plotted in Fig. \ref{figure1XXZ_25}, in that such a density only vanishes for $\Delta \rightarrow 1$.

That the excitation-energy density $\delta E/N J$ is finite for $\Delta > 1$ is indeed a universal property that applies 
to all states of class (B) of the spin-${1\over 2}$ XXZ chain at $h=0$. 

Our results apply to {\it all} finite temperatures $T>0$ and are consistent with those of Ref. \cite{Carmelo_25}, the
studies of which found that for $\Delta >1$ the spin diffusion constant is finite both for very small temperatures and very high temperatures. 

The expected normal diffusive spin transport for all finite temperatures when $\Delta >1$ \cite{Carmelo_25,Ljubotina_17}
remained though an unsolved issue. It is clarified in this paper
as a result of the identification of mechanisms that control the type of $T>0$ spin transport.

\section{Finite-temperature charge transport in the 1D gapped lattice spinless fermion model}
\label{SECV}

The Hamiltonian, Eq. (\ref{HD1}), and current operator, Eq. (\ref{s-current}), can be expressed in terms of spinless fermion operators. To
do so, we consider the inverse of the Jordan-Wigner transformation \cite{Jordan_28}, which for the spin-${1\over 2}$ XXZ chain at $h=0$ and
$\Delta \geq 1$ is given by,
\begin{eqnarray}
S_j^- & = & e^{-i\pi\sum_{j'=1}^{j-1}c_{j'}^{\dag}\,c_{i'}} c_{j}
\nonumber \\
S_j^+ & = & c_{j}^{\dag}e^{i\pi\sum_{j'=1}^{j-1}c_{j'}^{\dag}\,c_{i'}} \, ,
\label{InJW}
\end{eqnarray}
for $j= 1,...,N$. The spinless fermion operators $c_{j}^{\dag}$ and $c_{j}$ obey the usual anticommutation relations.
The operator $c_{j}^{\dag}$ creates a spinless fermion at site $j$, yet such a fermion is highly non-local.

The use of the relations, Eq. (\ref{InJW}), in the Hamiltonian, Eq. (\ref{HD1}),
of the spin-${1\over 2}$ XXZ chain at $h=0$, gives the following Hamiltonian for the
1D lattice spinless fermion model at zero chemical potential, $\mu =0$, 
\begin{equation}
\hat{H}  = -{J\over 2}\sum_{\langle j,i \rangle}c_{j}^{\dag}\,c_{i} + 
V\sum_{j=1}^N\left(\hat{n}_{j} - {1\over 2}\right) \left(\hat{n}_{j+1} - {1\over 2}\right) \, ,
\label{H}
\end{equation}
the lattice of which has $j= 1,...,N$ sites. Here $\sum_{\langle j,i \rangle}$ means summation 
over nearest neighbors, $\hat{n}_{j} = c_{j}^{\dag}\,c_{j}$ is the usual local number
operator, $J/2 = t$ is the hopping integral $t$, and $V$ is the nearest-neighbor Coulomb repulsion.

Our results refer to $V/J \geq 1$ and $\mu =0$. The 1D lattice spinless fermion model at $\mu =0$
is gapped for $V/J >1$. Its $\mu =0$ ground state has one fermion for each two lattice sites.
The hopping integral of the Hamiltonian, Eq. (\ref{H}), is usually called $t$, but in its expression we 
called it $J/2$ to emphasize the important exact correspondence between this model and the gapped spin-${1\over 2}$ XXZ chain,
Eq. (\ref{HD1}).

In the case of the spin current operator, Eq. (\ref{s-current}), the use of the relations, Eq. (\ref{InJW}), gives
the following charge current operator,
\begin{equation}
\hat{J}_c = i\,{J\over 2}\sum_{j=1}^{N}(c_{j}^{\dag}c_{j+1} - c_{j+1}^{\dag}c_{j})  \, .
\label{c-current}
\end{equation}

The transformation, Eq. (\ref{InJW}), maps the spin anisotropy $\Delta$ on the
ratio $V/J = \cosh \eta$ and is such that the Hamiltonian, Eq. (\ref{H}), is the Hamiltonian, Eq. (\ref{HD1}), 
expressed in terms of spinless fermion operators. Also the current operator, Eq. (\ref{c-current}), is that
given in Eq. (\ref{s-current}) expressed in terms of spinless fermion operators. The
only difference is that the former is now associated with charge degrees of freedom.

From the interplay of the 1D lattice spinless fermion model's Bethe-ansatz solution \cite{Peres_99,Gu_02} and symmetry, we find 
that the numbers of spinless fermions and spinless holes can be written as $N_f = {\cal{N}}_f + {\cal{M}}_f$
and $N_f^h = N - N_f = {\cal{N}}_f^h + {\cal{M}}_f^h$, respectively. This refers to
two types of spinless fermions and spinless holes: (i) a number ${\cal{N}}_f = N_f - \sum_{n=1}^{\infty}n N_n$
of unpaired spinless fermions and ${\cal{N}}_f^h = N_f^h - \sum_{n=1}^{\infty}n N_n$ of unpaired spinless holes 
and (ii) a number ${\cal{M}}_f = \sum_{n=1}^{\infty}n N_n$ of paired spinless fermions and
${\cal{M}}_f^h = \sum_{n=1}^{\infty}n N_n$ of paired spinless holes. 

Each pair involves one spinless fermion
and one spinless hole. We call $n$-pair a configuration of $n>1$ bound such pairs. For $n=1$, 
a $n$-pair refers to a single pair. $N_n$ is the number of $n$-pairs of an energy eigenstate.

Such numbers obey the sum-rule, 
\begin{equation}
N = {\cal{N}}_f + {\cal{N}}_f^h + \sum_{n=1}^{\infty}2n N_n \, .
\label{N}
\end{equation}

There is a Bethe ansatz for (i) $N_f \leq N/2$ and (ii) $N_f \geq N/2$ for which
(i) ${\cal{N}}_f = 0$ and ${\cal{N}}_f^h = N - \sum_{n=1}^{\infty}2n N_n$ and (ii) ${\cal{N}}_f = N - \sum_{n=1}^{\infty}2n N_n$ 
and ${\cal{N}}_f^h = 0$, respectively. For energy eigenstates outside the Bethe-ansatz subspace, one has that 
both ${\cal{N}}_f$ and ${\cal{N}}_f^h$ are finite and such that ${\cal{N}}_f + {\cal{N}}_f^h = N - \sum_{n=1}^{\infty}2n N_n$.

The Bethe-ansatz solution of the lattice 1D spinless fermion model 
involves one $n$-band for each $n=1,...,\infty$ branch of $n$-pairs with fixed value $n$.
The corresponding discrete $n$-band momentum values $q_j  = {2\pi\over N}\,I_j^n + q_n^{*}$ are such that $q_{j+1}-q_j = {2\pi\over N}$.
Consistently with the relation to the spin-${1\over 2}$ XXZ chain, the momentum $q_n^{*}$ appearing 
here vanishes for $V/J \rightarrow 1$ and for $V/J >1$ vanishes in the thermodynamic 
limit for ground states and excited states generated from them by a finite number of $n$-band processes.

For $N$ even the quantum numbers $I_j^n$ are given by,
\begin{eqnarray}
I_j^n & = & \pm 1/2, \pm 3/2, ..., \pm {L_n -1\over 2} \hspace{0.20cm}{\rm for}\hspace{0.20cm}N_n\hspace{0.20cm}{\rm even}
\nonumber \\
& = & 0, \pm 1, \pm 2, ..., \pm {L_n -1\over 2} \hspace{0.20cm}{\rm for}\hspace{0.20cm}N_n\hspace{0.20cm}{\rm odd} \, .
\label{Ijn}
\end{eqnarray}
Here $j = 1,...,L_n$, $L_n = N_n + N_n^h$, and the number $N_n^h$ of $n$-holes is given by,
\begin{equation}
N_n^h = {\cal{N}}_f^h + {\cal{N}}_f + \sum_{n'=n+1}^{\infty}2(n' - n)N_{n'} \, .
\end{equation}
The $n$-band shifted discrete momentum values $q_j - q_n^{*}$ have intervals 
$q_j - q_n^{*} \in [q_n^-,q_n^+]$ where $q_n^{\pm} = \pm {\pi\over N}(L_n -1)$.

We write the number $N_f$ of spinless fermions and $N_f^h = N - N_f$ of spinless holes as,
\begin{eqnarray}
N_f & = & N/2 - R_f
\nonumber \\
N_f^h & = & N - N_f = N/2 + R_f \hspace{0.20cm}{\rm where}
\nonumber \\
R_f & = & {1\over 2}({\cal{N}}_f^h - {\cal{N}}_f) \in [-S_f,S_f]
\nonumber \\
S_f & = & {1\over 2}({\cal{N}}_f^h + {\cal{N}}_f) \in [0,N/2]\, .
\end{eqnarray}
The two intervals $R_f \in [-S_f,0]$ and $R_f \in [0,S_f,]$ refer to the two alternative descriptions of charge transport in 
terms of spinless-fermion charge carriers and spinless-hole 
charge carriers, respectively. The transformation, Eq. (\ref{InJW}), maps $S^z$ on $R_f = {1\over 2}({\cal{N}}_f^h - {\cal{N}}_f) $ and $S_q$
on $S_f = {1\over 2}({\cal{N}}_f^h + {\cal{N}}_f)$.

The 1D lattice spinless fermion model is solvable by Bethe anstaz in the presence of a uniform vector potential $\Phi$
(twisted boundary conditions) \cite{Peres_99,Gu_02}. Both the $N_f$ spinless fermions and the $N_f^h = N - N_f$ spinless 
holes couple to the vector potential, their couplings having opposite sign. 

It follows that the coupling of the ${\cal{M}}_f = \sum_{n=1}^{\infty}n N_n$ paired spinless fermions and
${\cal{M}}_f^h = \sum_{n=1}^{\infty}n N_n$ paired spinless holes vanishes. Indeed, the couplings of the spinless fermion and
spinless hole in each pair cancel each other. Therefore, only the number ${\cal{N}}_f$ of unpaired spinless fermions and the 
number ${\cal{N}}_f^h$ of unpaired spinless holes couple to the vector potential and are thus the charge carriers.

For the energy eigenstates outside the Bethe-ansatz subspace, we have that $-S_f < R_f < S_f$. There are two Bethe ansatz
for which $R_f = -S_f$ and $R_f = S_f$, respectively. We choose that for which $R_f = S_f$ and
denote an energy eigenstate inside that Bethe-ansatz subspace by $\left\vert \iota^{\eta}, S_f,S_f\right\rangle$.
Here $\iota^{\eta}$ are all quantum numbers needed to specify the state other than $S_f$. 
We can then generate from such a state a number $2S_f$ of states outside the $R_f = S_f$ Bethe ansatz as,
\begin{equation} 
\left\vert \iota^{\eta}, S_f, R_f\right\rangle 
= {1\over \sqrt{{\cal{C}}_f}}\Bigl(\sum_{j=1}^N c_{j}^{\dag}\,e^{i\pi\sum_{j'=1}^{j-1}c_{j'}^{\dag}\,c_{i'}}\Bigr)^{n_f}
\left\vert \iota^{\eta}, S_f,S_f\right\rangle \, .
\label{state}
\end{equation} 
Here $n_f = S_f + R_f = 0,1, ... ,2S_f$ for $R_f \in [-S_q,S_q]$ and,
\begin{equation}
{\cal{C}}_{f} = 
\prod_{l=1}^{n_f}{\sinh^2 (\eta\,(S_f +1/2)) - \sinh^2 (\eta\,(l - S_f - 1/2))\over\sinh^2 \eta} \, ,
\label{nonBAstatesDelta1}
\end{equation}
where $\eta$ is such that $\cosh\eta = V/J$. For $n_f = 0,1, ... ,S_f - 1$ the charge carriers are spinless holes
and for $n_f = S_f+1, ... ,2S_f$ the charge carriers are spinless fermions.

What for the spin-${1\over 2}$ XXZ chain for $h=0$ and $\Delta \geq 1$ is a generator of spin flips
is in Eq. (\ref{state}) for the 1D lattice spinless fermion model
for $\mu =0$ and $V/J \geq 1$ the generator in Eq. (\ref{state})
that creates a number $n_f$ of spinless fermions. 

For the latter model the charge elementary currents carried by the charge
carriers also play an important role in $T>0$ charge transport. They are given by,
\begin{eqnarray}
j_c (S_f) & = & {\langle \iota^{\eta},S_f,S_f\vert\hat{J}_c\vert \iota^{\eta}, S_f,S_f\rangle \over 2S_f} 
\nonumber \\
& = & {\langle \iota^{\eta},S_f,S_f\vert\hat{J}_c\vert \iota^{\eta}, S_f,S_f\rangle \over {\cal{N}}_f^h + {\cal{N}}_f} \, ,
\label{jc}
\end{eqnarray}
where $\hat{J}_c$ is the charge current operator, Eq. (\ref{c-current}). This charge elementary current is
carried by one spinless hole. The use of the alternative Bethe ansatz gives a charge elementary current
given by $\langle \iota^{\eta},S_f,-S_f\vert\hat{J}_c\vert \iota^{\eta}, S_f,-S_f\rangle/2S_f = - j_c (S_f)$, which
is carried by one spinless fermion.

The charge current expectation value of any $S_f >0$ energy eigenstate
$\vert \iota^{\eta}, S_f,R_f\rangle$, Eq. (\ref{state}), can 
be expressed in terms of the charge elementary current $j_c (R_f)$, Eq. (\ref{jc}), as,
\begin{equation}
\langle \iota^{\eta}, S_f,R_f \vert\hat{J}_c\vert \iota^{\eta}, S_f,R_f\rangle
= 2R_f\times j_c (S_f) \, ,
\label{expecsate}
\end{equation}
where $R_f \in [-S_f,S_f]$. Such a charge current expectation value has opposite sign when the 
charge carriers are spinless holes for $R_f \in [1,S_f]$ and spinless fermions for $R_f \in [-S_f,-1]$

By the use of manipulations of the Kubo formula and Einstein relation similar to those
made for other 1D correlated models \cite{Carmelo_25,Carmelo_24,Carmelo_25B,Carmelo_25A},
we find that the charge-diffusion constant $D_c (T)$ associated with the regular part of the 
charge conductivity $\sigma_{c,{\rm reg}} (\omega,T)$, Eq. (\ref{sigma}) for $\alpha =c$, can
be expressed in terms of the charge elementary currents $j_c (S_f)$, Eq. (\ref{jc}), as,
\begin{equation}
D_c (T) = C_c (T)\,\Pi_c (T) = C_c (T)\,N\,\langle\vert j_c (S_f)\vert^2\rangle_{T} \, ,
\label{DproptoOmegac}
\end{equation}
where the thermal expectation value is given by,
\begin{eqnarray}
&& \langle\vert j_c (R_f)\vert^2\rangle_{T} = {\Pi_c (T)\over N} 
\nonumber \\
&& = \sum_{S_f=1}^{N/2}\sum_{R_f=-S_f}^{S_f}\sum_{\iota^{\eta}} 
p_{\iota^{\eta}, S_f,R_f }\vert j_c (R_f)\vert^2 \, ,
\label{jz2TDc}
\end{eqnarray}
and $p_{\iota^{\eta}, S_f,R_f }$ are the Boltzmann weights.
We have here excluded $S_f=0$ in the summation $\sum_{S_f=1}^{N/2}$ because $S_f=0$ energy eigenstates have
zero charge current expectation value.

The coefficient $C_c (T) = 1/[8 v_{c,{\rm LR}}\,\chi_c (T)\,f_{1} (T)]$ in Eq. (\ref{DproptoOmegac}) is finite for $T>0$. 
In its expression $v_{c,{\rm LR}}$ and $\chi_c (T)$ are the charge Lieb-Robinson velocity and static charge susceptibility, respectively, 
and $f_{1} (T)$ is the second derivative of the free-energy density with respect to 
$2R_f/N$ at $2R_f/N = 0$. 

The $S_f >0$ energy eigenstates are again of two classes, (A) and (B), the square of the charge elementary currents
$\vert j_c (R_f)\vert^2$ of which are of order of $1/N$ and finite, respectively. The former vanish in the thermodynamic
limit, $N\rightarrow\infty$.

By the use of the same procedures as for the spin-${1\over 2}$
XXZ chain at $h=0$ for $\Delta \geq 1$, we have studied the dependence on $V/J$ of the excitation-energy density
of the states of class (B) of the 1D lattice spinless fermion model at $\mu =0$ for $V/J \geq 1$. 

Consistently with according to the inverse of the Jordan-Wigner transformation, Eq. (\ref{InJW}),
the Hamiltonian, Eq. (\ref{HD1}), of the spin-${1\over 2}$ XXZ chain at $h=0$ being mapped on the
Hamiltonian of the 1D lattice spinless fermion model at $\mu =0$, Eq. (\ref{H}), and the anisotropy 
$\Delta$ being mapped on the ratio $V/J$, it is found that for the latter model
the excitation-energy density of the states of class (B)
vanishes at $V/J =1$ and is finite for $V/J >1$. 

Using the same arguments as for the spin-${1\over 2}$ XXZ chain at $h=0$, which account for the form of the 
charge-diffusion constant, Eq. (\ref{DproptoOmegac}), this then implies that for the 1D lattice spinless 
fermion model at $\mu =0$, Eq. (\ref{H}), $T>0$ charge transport is anomalous 
superdiffusive at $V/J = 1$ and normal diffusive for $V/J >1$.

In contrast to the 1D Hubbard model at $\mu=0$, the $T>0$ charge transport of which is always normal diffusive 
for onsite repulsion $U>0$ \cite{Carmelo_24,Carmelo_25B,Carmelo_25A}, the the 1D lattice spinless fermion 
model at $\mu=0$ has a point, $V/J = 1$, at which its $T>0$ charge transport is anomalous superdiffusive.

\section{Concluding remarks}
\label{SECVI}

The studies of this paper are on $T>0$ spin transport in the spin-${1\over 2}$ XXZ chain at $h=0$,
Eq. (\ref{HD1}), for anisotropy $\Delta \geq 1$ and $T>0$ charge transport in the 1D lattice spinless fermion model at $\mu=0$,
Eq. (\ref{H}), for $V/J\geq 1$. Our detailed studies focused on the $T>0$ spin transport in the
gapped spin-${1\over 2}$ XXZ chain at $h=0$. We then combined the Jordan-Wigner transformation that connects both models 
with the Bethe-ansatz solution of the 1D lattice spinless fermion model to
reach important information on the $T>0$ charge transport for the latter model.

Concerning the spin-${1\over 2}$ XXZ chain at $h=0$, we have accounted for the contributions to the
spin diffusion constant for anisotropy $\Delta \geq 1$
from $S_q>0$ energy eigenstates of class (A) and (B), respectively, the spin elementary
currents of which carried by a number $2S_q$ of spin carriers is of order $1/N$
and finite, respectively. The former elementary currents vanish in the
thermodynamic limit, $N\rightarrow \infty$. 
 
The states of class (A) are then found to provide finite contributions to the
spin diffusion constant for $T>0$, $D_s (T) = C_s (T)\,N\,\langle\vert j_{s,\pm 1/2}\vert^2\rangle_{T}$, Eq. (\ref{DproptoOmega}),
where the coefficient $C_s (T)$ is finite. In contrast, in the thermodynamic limit the finite density of states of class (B) provides
infinite contributions to the spin diffusion constant at $k_B T$ values corresponding to
the excitation energies associated with their energy eigenvalues.

It was found in this paper that such excitation energies are infinite for $\Delta >1$ and finite at $\Delta =1$.
In Fig. \ref{figure1XXZ_25} this refers to the corresponding excitation-energy densities being
finite and vanishing, respectively. Hence  for $\Delta >1$ there are no infinite contributions to the 
spin diffusion constant, Eq. (\ref{DproptoOmega}), from the states of class (B) at finite values of $k_B T < \infty$.
Since the states of class (A) provide finite contributions to that diffusion constant, it
is finite for $\Delta >1$ and all finite temperatures $T>0$. 

This mechanism confirms the expectation \cite{Carmelo_25,Ljubotina_17} that 
spin transport is normal diffusive for the spin-${1\over 2}$ XXZ chain at $h=0$
for $\Delta >1$ and all finite temperatures $T>0$.

On the other hand, at $\Delta =1$ the states of class (B) provide infinite contributions
to the spin diffusion constant at all finite values of $k_B T$ because their excitation energies 
 $\delta E \in ]0,\infty]$ associated with their energy eigenvalues are finite. This implies anomalous
superdiffusive spin transport at all finite temperatures for the isotropic spin-${1\over 2}$ XXX chain at $h=0$,
as found by hydrodynamic theory and KPZ scaling  \cite{Ljubotina_19,Nardis_20,Nardis_23}. 

By combining the inverse of the Jordan-Wigner transformation, Eq. (\ref{InJW}), 
that connects the spin $1/2$ XXZ chain at $h=0$ for $\Delta \geq 1$ to the 1D lattice spinless fermion model
at $\mu =0$ for $V/J \geq 1$ with the Bethe-ansatz solution of the latter model,
we found that its $T>0$ charge transport is anomalous superdiffusive at $V/J = 1$ and normal diffusive for $V/J >1$.
In both models, $T>0$ anomalous superdiffusive transport and normal diffusive transport occurs when
the spectrum is gapless and gapped, respectively.

The dominant $T>0$ spin transport in the 
spin-${1\over 2}$ XXZ chain is in the presence of a finite magnetic
field, $h>0$, ballistic rather than diffusive or superdiffusive. This is due to the spin stiffness being finite,
vanishing as $h^2$ for $h\sim 0$. (See for instance Eq. (6) of Ref. \onlinecite{Ilievski_18}, which applies to several
1D quantum lattice models, including the spin $1/2$ XXZ chain for $\Delta\geq 1$.) 

In what concerns experimental systems and measurement techniques where the predicted
superdiffusive and diffusive behaviors could be observed, a strong manifestation of integrable dynamics 
was recently found in the 1D Heisenberg magnet, KCuF$_3$: Its neutron scattering experiments showed
clear evidence of superdiffusive spin transport at temperatures exceeding 100 K \cite{Scheie_21}.
The spin dynamics in CaCu$_2$O$_3$ and SrCuO$_2$ is also believed to be governed by the antiferromagnetic 
spin-${1\over 2}$ Heisenberg chain \cite{Hess_19}. Hamiltonian-engineering methods on
nuclear spins have been used in the material fluorapatite \cite{Wei_19} in which both the isotropic 
spin-${1\over 2}$ XXX chain and the anisotropic spin-${1\over 2}$ XXZ chain can be realized. That for anisotropy $\Delta >1$ 
and $T>0$ the spin-${1\over 2}$ XXZ spin chain exhibits diffusive spin transport was recently verified in 
fluorapatite \cite{Peng_23}. 

All the above materials are expected to have very small yet finite second-neighbor interactions. Their
experimental results seem to indicate that such very small interactions may be masked by thermal 
vibrations or other effects and thus do not lead to significant deviations from the integrable spin-${1\over 2}$ XXZ chain. 
However, second- and third-neighbor interactions break integrability. For materials with increased values of such 
interactions it is expected that they significantly affect the $T>0$ spin transport behavior. 

Integrable 1D lattice systems have infinitely many conserved quantities that in the case of the
spin-${1\over 2}$ XXZ chain for $\Delta \geq 1$ are associated with the $n=1,...,\infty$ $n$-pairs.
When integrability is broken, only a few residual conserved quantities survive,
eventually leading to thermalization, chaotic dynamics, and conventional hydrodynamics \cite{Bastianello_21}.
This means that the $n$-pairs that play a key role in our study become illy defined and do not
refer to energy eigenstates of the nonintegrable system. The integrability breaking perturbations lead to force terms and to collision integrals, 
which rely on the matrix elements of generic local operators between eigenstates of the unperturbed integrable system.
Determining these matrix elements is a challenging task. 

The nonintegrable perturbations associated with second- and third-neighbor interactions have thus highly nonperturbative 
effects, and breaking integrability poses very profound questions beyond the goals of this paper.
Ultra-cold gases can also be used to 
realize integrable lattice models such as the spin-${1\over 2}$ XXZ chain. The symmetry
of the spin-spin interactions depends on the microscopic state-dependent 
scattering properties of the cold atoms. In general, it is approximately
isotropic, $\Delta\approx 1$. However, by tuning the system close to a state-dependent Feshbach 
resonance one can realize anisotropic spin-${1\over 2}$ XXZ chains with tunable anisotropy $\Delta$ \cite{Jepsen_20}.

Concerning the relation to other quantum problems, we consider the 1D Hubbard model,
that in metallic phases for $\mu >0$ and $h\geq 0$ is at low excitation energies conformal invariant 
\cite{Carmelo_93} and at higher excitation energies describes the dynamical properties of some metallic 
quasi-1D materials \cite{Carmelo_06}.

Here we refer to the recent results of Ref. \onlinecite{Carmelo_25A} on $T>0$ charge and spin transport
in the 1D Hubbard model at the $h = \mu =0$ point for onsite repulsion $U>0$.
Such studies use an exact representation in terms of
a number $L_s$ of physical spins and $L_{\eta}$ of physical $\eta$-spins the configurations 
of which generate the irreducible representations of the model's spin SU(2) symmetry and
$\eta$-spin SU(2) symmetry, respectively \cite{Carmelo_25B}. Here $L_s+L_{\eta} = L$ where $L$ denotes both the
lattice length and its number of sites. The number $L_{\eta}$ of physical $\eta$-spins are associated
with the charge degrees of freedom \cite{Carmelo_25B}. 

In the case of $T>0$ spin transport, the parameter $\Delta_{s\eta} = 2 + z\in [1,3]$ where $z = (L_{\eta}-L_s)/L \in [-1,1]$
of the 1D Hubbard model at the $h = \mu = 0$ point \cite{Carmelo_25A}
is found to play somehow the same role as anisotropy for the spin-${1\over 2}$ XXZ chain at $h=0$.
It is such that when $\Delta_{s\eta} = 1$ and thus $L_s = L$ and $L_{\eta} =0$, the
model describes a spin-only quantum problem that for large onsite repulsion $U$ is equivalent
to the spin-${1\over 2}$ XXX chain at $h=0$. For $1<\Delta_{s\eta}<3$ the quantum problem
has both spin and charge degrees of freedom that interact with each other.
Finally, when  $\Delta_{s\eta} = 3$ and thus $L_s = 0$ and $L_{\eta} = L$, the
model describes a charge-only quantum problem.

The excitation-energy density of this $\Delta_{s\eta} \in [1,3]$ quantum problem,
the charge and spin elementary currents of which carried by the corresponding carriers 
are finite, has for $T>0$ spin transport similarities to that of the spin-${1\over 2}$ XXZ chain at $h=0$
plotted in Fig. \ref{figure1XXZ_25}. 

For $U>0$ the excitation-energy density of the states
of class (B) is finite for $\Delta_{s\eta} >1$ and only vanishes at $\Delta_{s\eta} = 1$,
when the corresponding spin-only quantum problem has $T>0$ anomalous superdiffusive 
spin transport. For $\Delta_{s\eta} >1$ there emerge for $U>0$ increasing interactions
between the spin and charge degrees of freedom that the $L_s$ physical spins
feel as a kind of anisotropy.

The spin diffusion constant of the 1D Hubbard model at the $h=\mu=0$ point
has for $U>0$ contributions from summations over all $\Delta_{s\eta} \in [1,3]$ values. The contributions
from $\Delta_{s\eta} = 1$ justify why the $T>0$ spin transport is anomalous
superdiffusive for the 1D Hubbard model at the $h=\mu =0$ point for $U>0$, as also found by
hydrodynamic theory, KPZ scaling, and other methods \cite{Ilievski_18,Moca_23,Fava_20}.

On the other hand, that the excitation-energy density of the states of class (B) is finite for $\Delta_{s\eta} >1$ 
is shown in Ref. \onlinecite{Carmelo_25A} to imply that the $T>0$ charge transport is normal
diffusive for the 1D Hubbard model at the $h=\mu =0$ point for $U>0$, in contrast to predictions
of hydrodynamic theory and KPZ scaling. 

For the 1D Hubbard model at the $h=\mu =0$ point, $T>0$ charge transport is normal diffusive 
for onsite repulsion $U>0$ and becomes ballistic at $U=0$ \cite{Carmelo_25A}. This differs from
the 1D lattice spinless fermion model at $\mu=0$, Eq. (\ref{H}), that has $T>0$ normal diffusive
charge transport for $V/J>1$, as the 1D Hubbard model for $U>0$, yet it becomes $T>0$ anomalous superdiffusive 
charge transport at $V/J = 1$, rather than ballistic, as for the latter model at $U=0$.

Our results open the door to a key advance in the understanding of the $h=0$ 
spin transport for $T>0$ in the spin-${1\over 2}$ XXZ chain for anisotropy $\Delta \geq 1$ and
the $\mu =0$ charge transport for $T>0$ in the 1D lattice spinless fermion model for $V/J\geq 1$.

\acknowledgements
This work was supported by FCT under research unit UID/04540 - Center of Physics and Engineering of Advanced Materials and
contract LA/P/0095/2020, LaPMET, Laboratory of Physics for Materials and Emerging Technologies - and under
research unit UIDB/04650.\\ \\ 
\appendix

\section{Spin elementary currents and excitation-energy densities of class (B) states}
\label{A}

\subsection{$\bar{n}$ states}
\label{A1}

The $n$-band occupancies of the $\bar{n}$ states are given in Eq. (\ref{Nnstates}) in terms
of rapidity variables $\varphi \in [-\pi,\pi]$. Here we shortly discuss their $n$-band occupancies in terms 
of the $n$-band shifted momentum $q - q_n^{\Delta} \in [q_n^-,q_n^+]$ where,
\begin{equation}
q_n^{\pm} = q_n (\pm\pi) - q_n^{\Delta} = \pm {\pi\over N}(L_n -1) \, ,
\label{qqq}
\end{equation}
and $q_n^{\Delta}$ has limiting values given by, 
\begin{eqnarray}
q_n^{\Delta} & = & 0\hspace{0.20cm}{\rm for}\hspace{0.20cm}\Delta\rightarrow 1
\nonumber \\
& = & \sum_{n' =1}^{\infty} {n+n' - \vert n-n'\vert\over 2\pi}
\int_{-\pi}^{\pi}\,d\varphi\tilde{N}_{n'} (\varphi)\,2\pi\sigma_{n'} (\varphi)\,\varphi
\nonumber \\
& - & {1\over 2\pi}\int_{-\pi}^{\pi}\,d\varphi\tilde{N}_n (\varphi)\,2\pi\sigma_n (\varphi)\,\varphi 
\hspace{0.20cm}{\rm for}\hspace{0.20cm}\Delta\rightarrow\infty \, .
\label{qDelta}
\end{eqnarray}

The $n$-band occupancies of the $\bar{n}$ states given in Eq. (\ref{Nnstates}) in terms
of rapidity variables are under the transformation, Eq. (\ref{dqdvar}),
given by $N_{n} (q) = 0$ for $q-q_n^{\Delta} \in [q_n^-,q_n^0]$ and
$N_{n} (q) = 1$ for $q-q_n^{\Delta} \in [q_n^0,q_n^+]$ for $n=1,...,\bar{n}$
and $N_{n} (q) = 0$ for $n > \bar{n}$ in terms of $n$-band momentum values \cite{Carmelo_25}.
Here $q_n^0 = q_n (0)-q_n^{\Delta}\geq 0$ for $n=1,...,\bar{n}$.

The $n$-band momentum functions $q_n (\varphi)$, with boundary conditions 
$q_n (\pm\pi) = q_n^{\pm} + q_n^{\Delta}$, are for $\bar{n}$ states 
indeed such that $q_n^0 = q_n (0)-q_n^{\Delta}\geq 0$ for $n=1,...,\bar{n}$, the equality 
$q_n^0 = 0$ being reached only in the $\Delta\rightarrow\infty$ limit \cite{Carmelo_25}. 

This means that while for $\bar{n}$ states the occupancy of the $n=1,...,\bar{n}$ bands is half filled 
in terms of the rapidity variable $\varphi$, Eqs. (\ref{Nnstates}) and (\ref{jznnRAPbarn}), 
except for $\Delta\rightarrow\infty$ it is not half filled in terms of the $n$-band shifted momentum 
$q - q_n^{\Delta}$. 

\subsection{Other states of class (B)}
\label{A2}

We considered all different types of states of class (B) with finite occupancies
in $n$ bands for $n=1,...,\bar{n}$ where $\bar{n}=1,...,\infty$. For instance,
we considered an alternative type of class (B) states that have half-filled occupancies in terms
of the $n$-band shifted momentum $q - q_n^{\Delta}$ for $\Delta >1$. We call them $q\bar{n}$ states

Since $q_n^{\pm} = q_n (\pm\pi) - q_n^{\Delta} = \pm {\pi\over N}(L_n -1)$, Eq. (\ref{qqq}), in terms of 
such $n$-band shifted momentum values the half-filled occupancy of the $n=1,...,\bar{n}$ bands reads as,
\begin{eqnarray}
N_{n} (q) & = & 0 \hspace{0.20cm}{\rm for}\hspace{0.20cm}q - q_n^{\Delta} \in [q_n^{-},0]
 \hspace{0.20cm}{\rm and}\hspace{0.20cm}n=1,...,\bar{n}
\nonumber \\
& = & 1 \hspace{0.20cm}{\rm for}\hspace{0.20cm}q - q_n^{\Delta}\in [0,q_n^{+}]
\hspace{0.20cm}{\rm and}\hspace{0.20cm}n=1,...,\bar{n}
\nonumber \\
N_{n} (q) & = & 0  \hspace{0.20cm}{\rm for}\hspace{0.20cm}n > \bar{n} \, .
\label{Nnstatesq}
\end{eqnarray}
The $q\bar{n}$ states have been inherently constructed to the $n = 1,...,\bar{n}$ momentum bands having 
for shifted momentum values $q - q_n^{\Delta}$ half-filled occupancy for $\Delta >1$.

In terms of $n$-band momentum hole distributions $N_n^h (q)$, Eq. (\ref{dqdvarholes}),
the spin elementary currents of the class (B) $q\bar{n}$ states reads.
\begin{eqnarray}
&& j_{s,\pm 1/2} (\bar{n}) = \mp {J\sinh\eta\over\pi}\sum_{n=1}^{\bar{n}}{N\over N_n^h}
\nonumber \\
&& \times \int_{q_n^- + q_n^{\Delta}}^{q_n^{\Delta}}{dq\over 2\pi\sigma_n (\varphi_n (q)}\,{n \sinh (n\eta)\sin \varphi_n (q)\over (\cosh (n\eta) - \cos \varphi_n (q))^2} \, ,
\nonumber \\
\label{jznnRAPbarn2}
\end{eqnarray}
where accounting for that $N_n = N_n^h$ for these states, it follows from Eq. (\ref{N}) that,
\begin{eqnarray}
{N\over N_n^h} & = & {N\over N_n}  = 1 + {\sum_{n'=1}^{\bar{n}}2n' N_{n'}\over 2S_q} \hspace{0.20cm}{\rm for}\hspace{0.20cm}n = \bar{n}
\nonumber \\
& = & {2S_q + \sum_{n'=1}^{\bar{n}}2n' N_{n'}\over 
N_n}\hspace{0.20cm}{\rm for}\hspace{0.20cm}n < \bar{n} \, .
 \nonumber \\
\end{eqnarray}
In contrast to the $\bar{n}$ states, for the $q\bar{n}$ states the ratio $N_n/N_n^h=1$ is independent of $\Delta$.

Except for $\Delta\rightarrow\infty$, the spin elementary currents, Eq. (\ref{jznnRAPbarn2}), 
carried by a number $2S_q$ of spin carriers that populate the $S_q>0$ $q\bar{n}$ states have absolute values smaller than those of the 
$S_q>0$ $\bar{n}$ states, Eq. (\ref{jznnRAPbarn}).
We have also computed the excitation-energy density in units of $J$ for the $q\bar{n}$ states, which reads,
\begin{eqnarray}
&& {\delta E\over N J} = \sinh\eta\,\Bigl({1\over 2} + 2 \sum_{l = 1}^{\infty}{1\over 1 + e^{2l\eta}}
\nonumber \\
&& - \sum_{n=1}^{\bar{n}}{1\over 2\pi}\int_{q_n^{\Delta}}^{q_n^{+}+q_n^{\Delta}}dq\,
{\sinh (n\,\eta)\over \cosh (n\,\eta) - \cos\varphi_n (q)}\Bigr) \, .
\label{Energyq}
\end{eqnarray}
Again, this excitation-energy density is always finite for $\Delta >1$ and only vanishes 
for $\Delta \rightarrow 1$, similarly to the spectra of the $S_q>0$ $\bar{n}$ states shown in
Fig. \ref{figure1XXZ_25}.

We also considered states of class (B) with rapidity occupancies similar to those given in Eq. (\ref{Nnstates}) 
but with the $\varphi$ values separating $n$-holes from $n$-pairs deviating a bit from
$\varphi = 0$. Again, such states have spin elementary currents smaller than those of the 
$\bar{n}$ states, Eq. (\ref{jznnRAPbarn}). The same applies to states of class (B) with $n$-band momentum $q$ occupancies similar 
to those given in Eq. (\ref{Nnstatesq}) but with the $n$-band shifted momentum value $q - q_n^{\Delta}$ separating $n$-holes from $n$-pairs 
deviating a bit from $q - q_n^{\Delta} = 0$.

Importantly, all such states have excitation-energy density spectra in function of $\Delta$ similar to those 
plotted in Fig. \ref{figure1XXZ_25}, as they only vanish for $\Delta\rightarrow 1$.

Finally, the functions $2\pi\sigma_{n} (\varphi)$ appearing in Eqs. (\ref{qDelta})
and (\ref{jznnRAPbarn2}) are solutions of the coupled integral equations given
in Eqs. (A3)-(A5) of Ref. \onlinecite{Carmelo_25}.

\end{document}